\documentstyle[mprocl]{article}
\def\fileversion{v1.20a}
\def\filedate{21.6.94}
\edef\epsfigRestoreAt{\catcode`@=\number\catcode`@\relax}%
\catcode`\@=11\relax
\ifx\undefined\@makeother                
\def\@makeother#1{\catcode`#1=12\relax}  
\fi                                      
\immediate\write16{Document style option `epsfig', \fileversion\space
<\filedate> (edited by SPQR + pks)}
\newcount\EPS@Height \newcount\EPS@Width \newcount\EPS@xscale
\newcount\EPS@yscale
\def\psfigdriver#1{%
  \bgroup\edef\next{\def\noexpand\tempa{#1}}%
    \uppercase\expandafter{\next}%
    \def\LN{DVITOLN03}%
    \def\DVItoPS{DVITOPS}%
    \def\DVIPS{DVIPS}%
    \def\emTeX{EMTEX}%
    \def\OzTeX{OZTEX}%
    \def\Textures{TEXTURES}%
    \global\chardef\fig@driver=0
    \ifx\tempa\LN
        \global\chardef\fig@driver=0\fi
    \ifx\tempa\DVItoPS
        \global\chardef\fig@driver=1\fi
    \ifx\tempa\DVIPS
        \global\chardef\fig@driver=2\fi
    \ifx\tempa\emTeX
        \global\chardef\fig@driver=3\fi
    \ifx\tempa\OzTeX
        \global\chardef\fig@driver=4\fi
    \ifx\tempa\Textures
        \global\chardef\fig@driver=5\fi
  \egroup
\def\psfig@start{}%
\def\psfig@end{}%
\def\epsfig@gofer{}%
\ifcase\fig@driver
\typeout{WARNING! ****
 no specials for LN03 psfig}%
\or 
\def\psfig@start{}%
\def\psfig@end{\special{dvitops: import \@p@sfilefinal \space
\@p@swidth sp \space \@p@sheight sp \space fill}%
\if@clip \typeout{Clipping not supported}\fi
\if@angle \typeout{Rotating not supported}\fi
}%
\let\epsfig@gofer\psfig@end
\or 
\def\psfig@start{\special{ps::[begin]  \@p@swidth \space \@p@sheight \space%
        \@p@sbbllx \space \@p@sbblly \space%
        \@p@sbburx \space \@p@sbbury \space%
        startTexFig \space }%
        \if@clip
                \if@verbose
                        \typeout{(clipped to BB) }%
                \fi
                \special{ps:: doclip \space }%
        \fi
        \if@angle              
                \special {ps:: \@p@sangle \space rotate \space}
        \fi
        \special{ps: plotfile \@p@sfilefinal \space }%
        \special{ps::[end] endTexFig \space }%
}%
\def\psfig@end{}%
\def\epsfig@gofer{\if@clip
                        \if@verbose
                           \typeout{(clipped to BB)}%
                        \fi
                        \epsfclipon
                  \fi
                  \epsfsetgraph{\@p@sfilefinal}%
}%
\or 
\typeout{WARNING. You must have a .bb info file with the Bounding Box
  of the pcx file}%
\def\psfig@start{}%
\def\psfig@end{\typeout{pcx import of \@p@sfilefinal}%
\if@clip \typeout{Clipping not supported}\fi
\if@angle \typeout{Rotating not supported}\fi
\raisebox{\@p@srheight sp}{\special{em: graph \@p@sfilefinal}}}%
\def\epsfig@gofer{}%
\or 
\def\psfig@start{}%
\def\psfig@end{%
\EPS@Width\@p@swidth
\EPS@Height\@p@sheight
\divide\EPS@Width by 65781  
\divide\EPS@Height by 65781
\special{epsf=\@p@sfilefinal
\space
width=\the\EPS@Width
\space
height=\the\EPS@Height
}%
\if@clip \typeout{Clipping not supported}\fi
\if@angle \typeout{Rotating not supported}\fi
}%
\let\epsfig@gofer\psfig@end
\or 
\def\psfig@end{
         \EPS@Width=\@bbw  
         \divide\EPS@Width by 1000
         \EPS@xscale=\@p@swidth \divide \EPS@xscale by \EPS@Width
         \EPS@Height=\@bbh  
         \divide\EPS@Height by 1000
         \EPS@yscale=\@p@sheight \divide \EPS@yscale by\EPS@Height
  \ifnum\EPS@xscale>\EPS@yscale\EPS@xscale=\EPS@yscale\fi
\if@clip
   \if@verbose
      \typeout{(clipped to BB)}%
   \fi
   \epsfclipon
\fi
\special{illustration \@p@sfilefinal\space scaled \the\EPS@xscale}%
}%
\def\psfig@start{}%
\let\epsfig\psfig
\else
\typeout{WARNING. *** unknown  driver - no psfig}%
\fi
}%
\newdimen\ps@dimcent
\ifx\undefined\fbox
\newdimen\fboxrule
\newdimen\fboxsep
\newdimen\ps@tempdima
\newbox\ps@tempboxa
\long\def\fbox#1{\leavevmode\setbox\ps@tempboxa\hbox{#1}\ps@tempdima\fboxrule
    \advance\ps@tempdima \fboxsep \advance\ps@tempdima \dp\ps@tempboxa
   \hbox{\lower \ps@tempdima\hbox
  {\vbox{\hrule height \fboxrule
          \hbox{\vrule width \fboxrule \hskip\fboxsep
          \vbox{\vskip\fboxsep \box\ps@tempboxa\vskip\fboxsep}\hskip
                 \fboxsep\vrule width \fboxrule}%
                 \hrule height \fboxrule}}}}%
\fi
\ifx\@ifundefined\undefined
\long\def\@ifundefined#1#2#3{\expandafter\ifx\csname
  #1\endcsname\relax#2\else#3\fi}%
\fi
\@ifundefined{typeout}%
{\gdef\typeout#1{\immediate\write\sixt@@n{#1}}}%
{\relax}%
\@ifundefined{epsfig}{}{\typeout{EPSFIG --- already loaded}\endinput}%
\@ifundefined{epsfbox}{\input epsf}{}%
\ifx\undefined\@latexerr
        \newlinechar`\^^J
        \def\@spaces{\space\space\space\space}%
        \def\@latexerr#1#2{%
        \edef\@tempc{#2}\expandafter\errhelp\expandafter{\@tempc}%
        \typeout{Error. \space see a manual for explanation.^^J
         \space\@spaces\@spaces\@spaces Type \space H <return> \space for
         immediate help.}\errmessage{#1}}%
\fi
\def\@whattodo{You tried to include a PostScript figure which
cannot be found^^JIf you press return to carry on anyway,^^J
The failed name will be printed in place of the figure.^^J
or type X to quit}%
\def\@whattodobb{You tried to include a PostScript figure which
has no^^Jbounding box, and you supplied none.^^J
If you press return to carry on anyway,^^J
The failed name will be printed in place of the figure.^^J
or type X to quit}%
\def\@nnil{\@nil}%
\def\@empty{}%
\def\@psdonoop#1\@@#2#3{}%
\def\@psdo#1:=#2\do#3{\edef\@psdotmp{#2}\ifx\@psdotmp\@empty \else
    \expandafter\@psdoloop#2,\@nil,\@nil\@@#1{#3}\fi}%
\def\@psdoloop#1,#2,#3\@@#4#5{\def#4{#1}\ifx #4\@nnil \else
       #5\def#4{#2}\ifx #4\@nnil \else#5\@ipsdoloop #3\@@#4{#5}\fi\fi}%
\def\@ipsdoloop#1,#2\@@#3#4{\def#3{#1}\ifx #3\@nnil
       \let\@nextwhile=\@psdonoop \else
      #4\relax\let\@nextwhile=\@ipsdoloop\fi\@nextwhile#2\@@#3{#4}}%
\def\@tpsdo#1:=#2\do#3{\xdef\@psdotmp{#2}\ifx\@psdotmp\@empty \else
    \@tpsdoloop#2\@nil\@nil\@@#1{#3}\fi}%
\def\@tpsdoloop#1#2\@@#3#4{\def#3{#1}\ifx #3\@nnil
       \let\@nextwhile=\@psdonoop \else
      #4\relax\let\@nextwhile=\@tpsdoloop\fi\@nextwhile#2\@@#3{#4}}%
\long\def\epsfaux#1#2:#3\\{\ifx#1\epsfpercent
   \def\testit{#2}\ifx\testit\epsfbblit
        \@atendfalse
        \epsf@atend #3 . \\%
        \if@atend
           \if@verbose
                \typeout{epsfig: found `(atend)'; continuing search}%
           \fi
        \else
                \epsfgrab #3 . . . \\%
                \epsffileokfalse\global\no@bbfalse
                \global\epsfbbfoundtrue
        \fi
   \fi\fi}%
\def\epsf@atendlit{(atend)}
\def\epsf@atend #1 #2 #3\\{%
   \def\epsf@tmp{#1}\ifx\epsf@tmp\empty
      \epsf@atend #2 #3 .\\\else
   \ifx\epsf@tmp\epsf@atendlit\@atendtrue\fi\fi}%

\chardef\trig@letter = 11
\chardef\other = 12
 
\newif\ifdebug 
\newif\ifc@mpute 
\newif\if@atend
\c@mputetrue 
 
\let\then = \relax
\def\r@dian{pt }%
\let\r@dians = \r@dian
\let\dimensionless@nit = \r@dian
\let\dimensionless@nits = \dimensionless@nit
\def\internal@nit{sp }%
\let\internal@nits = \internal@nit
\newif\ifstillc@nverging
\def \Mess@ge #1{\ifdebug \then \message {#1} \fi}%
 
{ 
        \catcode `\@ = \trig@letter
        \gdef \nodimen {\expandafter \n@dimen \the \dimen}%
        \gdef \term #1 #2 #3%
               {\edef \t@ {\the #1}
                \edef \t@@ {\expandafter \n@dimen \the #2\r@dian}%
                \t@rm {\t@} {\t@@} {#3}%
               }%
        \gdef \t@rm #1 #2 #3%
               {{%
                \count 0 = 0
                \dimen 0 = 1 \dimensionless@nit
                \dimen 2 = #2\relax
                \Mess@ge {Calculating term #1 of \nodimen 2}%
                \loop
                \ifnum  \count 0 < #1
                \then   \advance \count 0 by 1
                        \Mess@ge {Iteration \the \count 0 \space}%
                        \Multiply \dimen 0 by {\dimen 2}%
                        \Mess@ge {After multiplication, term = \nodimen 0}%
                        \Divide \dimen 0 by {\count 0}%
                        \Mess@ge {After division, term = \nodimen 0}%
                \repeat
                \Mess@ge {Final value for term #1 of
                                \nodimen 2 \space is \nodimen 0}%
                \xdef \Term {#3 = \nodimen 0 \r@dians}%
                \aftergroup \Term
               }}%
        \catcode `\p = \other
        \catcode `\t = \other
        \gdef \n@dimen #1pt{#1} 
}%
 
\def \Divide #1by #2{\divide #1 by #2} 
 
\def \Multiply #1by #2
       {{
        \count 0 = #1\relax
        \count 2 = #2\relax
        \count 4 = 65536
        \Mess@ge {Before scaling, count 0 = \the \count 0 \space and
                        count 2 = \the \count 2}%
        \ifnum  \count 0 > 32767 
        \then   \divide \count 0 by 4
                \divide \count 4 by 4
        \else   \ifnum  \count 0 < -32767
                \then   \divide \count 0 by 4
                        \divide \count 4 by 4
                \else
                \fi
        \fi
        \ifnum  \count 2 > 32767 
        \then   \divide \count 2 by 4
                \divide \count 4 by 4
        \else   \ifnum  \count 2 < -32767
                \then   \divide \count 2 by 4
                        \divide \count 4 by 4
                \else
                \fi
        \fi
        \multiply \count 0 by \count 2
        \divide \count 0 by \count 4
        \xdef \product {#1 = \the \count 0 \internal@nits}%
        \aftergroup \product
       }}%
 
\def\r@duce{\ifdim\dimen0 > 90\r@dian \then   
                \multiply\dimen0 by -1
                \advance\dimen0 by 180\r@dian
                \r@duce
            \else \ifdim\dimen0 < -90\r@dian \then  
                \advance\dimen0 by 360\r@dian
                \r@duce
                \fi
            \fi}%
 
\def\Sine#1%
       {{%
        \dimen 0 = #1 \r@dian
        \r@duce
        \ifdim\dimen0 = -90\r@dian \then
           \dimen4 = -1\r@dian
           \c@mputefalse
        \fi
        \ifdim\dimen0 = 90\r@dian \then
           \dimen4 = 1\r@dian
           \c@mputefalse
        \fi
        \ifdim\dimen0 = 0\r@dian \then
           \dimen4 = 0\r@dian
           \c@mputefalse
        \fi
        \ifc@mpute \then
                \divide\dimen0 by 180
                \dimen0=3.141592654\dimen0
                \dimen 2 = 3.1415926535897963\r@dian 
                \divide\dimen 2 by 2 
                \Mess@ge {Sin: calculating Sin of \nodimen 0}%
                \count 0 = 1 
                \dimen 2 = 1 \r@dian 
                \dimen 4 = 0 \r@dian 
                \loop
                        \ifnum  \dimen 2 = 0 
                        \then   \stillc@nvergingfalse
                        \else   \stillc@nvergingtrue
                        \fi
                        \ifstillc@nverging 
                        \then   \term {\count 0} {\dimen 0} {\dimen 2}%
                                \advance \count 0 by 2
                                \count 2 = \count 0
                                \divide \count 2 by 2
                                \ifodd  \count 2 
                                \then   \advance \dimen 4 by \dimen 2
                                \else   \advance \dimen 4 by -\dimen 2
                                \fi
                \repeat
        \fi
                        \xdef \sine {\nodimen 4}%
       }}%
 
\def\Cosine#1{\ifx\sine\UnDefined\edef\Savesine{\relax}\else
                             \edef\Savesine{\sine}\fi
        {\dimen0=#1\r@dian\multiply\dimen0 by -1
         \advance\dimen0 by 90\r@dian
         \Sine{\nodimen 0}%
         \xdef\cosine{\sine}%
         \xdef\sine{\Savesine}}}
\def\psdraft{\def\@psdraft{0}}%
\def\psfull{\def\@psdraft{1}}%
\psfull
\newif\if@compress
\def\pscompress{\@compresstrue}
\def\psnocompress{\@compressfalse}
\@compressfalse
\newif\if@scalefirst
\def\psscalefirst{\@scalefirsttrue}%
\def\psrotatefirst{\@scalefirstfalse}%
\psrotatefirst
\newif\if@draftbox
\def\psnodraftbox{\@draftboxfalse}%
\@draftboxtrue
\newif\if@noisy
\@noisyfalse
\newif\ifno@bb
\newif\if@bbllx
\newif\if@bblly
\newif\if@bburx
\newif\if@bbury
\newif\if@height
\newif\if@width
\newif\if@rheight
\newif\if@rwidth
\newif\if@angle
\newif\if@clip
\newif\if@verbose
\newif\if@prologfile
\def\@p@@sprolog#1{\@prologfiletrue\def\@prologfileval{#1}}%
\def\@p@@sclip#1{\@cliptrue}%
\newif\ifepsfig@dos  
\def\epsfigdos{\epsfig@dostrue}%
\epsfig@dosfalse
\newif\ifuse@psfig
\def\ParseName#1{\expandafter\@Parse#1}%
\def\@Parse#1.#2:{\gdef\BaseName{#1}\gdef\FileType{#2}}%

\def\@p@@sfile#1{%
  \ifepsfig@dos
     \ParseName{#1:}%
  \else
     \gdef\BaseName{#1}\gdef\FileType{}%
  \fi
  \def\@p@sfile{NO FILE: #1}%
  \def\@p@sfilefinal{NO FILE: #1}%
  \openin1=#1
  \ifeof1\closein1\openin1=\BaseName.bb
    \ifeof1\closein1
      \if@bbllx                 
        \if@bblly\if@bburx\if@bbury
          \def\@p@sfile{#1}%
          \def\@p@sfilefinal{#1}%
        \fi\fi\fi
      \else                     
        \@latexerr{ERROR. PostScript file #1 not found}\@whattodo
        \@p@@sbbllx{100bp}%
        \@p@@sbblly{100bp}%
        \@p@@sbburx{200bp}%
        \@p@@sbbury{200bp}%
        \psdraft
      \fi
    \else                       
      \closein1%
      \edef\@p@sfile{\BaseName.bb}%
      \typeout{using BB from \@p@sfile}%
      \ifnum\fig@driver=3
        \edef\@p@sfilefinal{\BaseName.pcx}%
      \else
        \ifepsfig@dos
          \edef\@p@sfilefinal{"`gunzip -c `texfind \BaseName.{z,Z,gz}"}%
        \else
          \edef\@p@sfilefinal{"`epsfig \if@compress-c \fi#1"}%
        \fi
      \fi
    \fi
  \else\closein1                
    \edef\@p@sfile{#1}%
    \if@compress  
      \edef\@p@sfilefinal{"`epsfig -c #1"}%
    \else
      \edef\@p@sfilefinal{#1}%
    \fi
  \fi%
}

\let\@p@@sfigure\@p@@sfile
\def\@p@@sbbllx#1{%
                                            \@bbllxtrue
                \ps@dimcent=#1
                \edef\@p@sbbllx{\number\ps@dimcent}%
                \divide\ps@dimcent by65536
                \global\edef\epsfllx{\number\ps@dimcent}%
}%
\def\@p@@sbblly#1{%
                \@bbllytrue
                \ps@dimcent=#1
                \edef\@p@sbblly{\number\ps@dimcent}%
                \divide\ps@dimcent by65536
                \global\edef\epsflly{\number\ps@dimcent}%
}%
\def\@p@@sbburx#1{%
                \@bburxtrue
                \ps@dimcent=#1
                \edef\@p@sbburx{\number\ps@dimcent}%
                \divide\ps@dimcent by65536
                \global\edef\epsfurx{\number\ps@dimcent}%
}%
\def\@p@@sbbury#1{%
                \@bburytrue
                \ps@dimcent=#1
                \edef\@p@sbbury{\number\ps@dimcent}%
                \divide\ps@dimcent by65536
                \global\edef\epsfury{\number\ps@dimcent}%
}%
\def\@p@@sheight#1{%
                \@heighttrue
                \global\epsfysize=#1
                \ps@dimcent=#1
                \edef\@p@sheight{\number\ps@dimcent}%
}%
\def\@p@@swidth#1{%
                \@widthtrue
                \global\epsfxsize=#1
                \ps@dimcent=#1
                \edef\@p@swidth{\number\ps@dimcent}%
}%
\def\@p@@srheight#1{%
                \@rheighttrue\use@psfigtrue
                \ps@dimcent=#1
                \edef\@p@srheight{\number\ps@dimcent}%
}%
\def\@p@@srwidth#1{%
                \@rwidthtrue\use@psfigtrue
                \ps@dimcent=#1
                \edef\@p@srwidth{\number\ps@dimcent}%
}%
\def\@p@@sangle#1{%
                \use@psfigtrue
                \@angletrue
                \edef\@p@sangle{#1}%
}%
\def\@p@@ssilent#1{%
                \@verbosefalse
}%
\def\@p@@snoisy#1{%
                \@verbosetrue
}%
\def\@cs@name#1{\csname #1\endcsname}%
\def\@setparms#1=#2,{\@cs@name{@p@@s#1}{#2}}%
\def\ps@init@parms{%
                \@bbllxfalse \@bbllyfalse
                \@bburxfalse \@bburyfalse
                \@heightfalse \@widthfalse
                \@rheightfalse \@rwidthfalse
                \def\@p@sbbllx{}\def\@p@sbblly{}%
                \def\@p@sbburx{}\def\@p@sbbury{}%
                \def\@p@sheight{}\def\@p@swidth{}%
                \def\@p@srheight{}\def\@p@srwidth{}%
                \def\@p@sangle{0}%
                \def\@p@sfile{}%
                \use@psfigfalse
                \@prologfilefalse
                \def\@sc{}%
                \if@noisy
                        \@verbosetrue
                \else
                        \@verbosefalse
                \fi
                \@clipfalse
}%
\def\parse@ps@parms#1{%
                \@psdo\@psfiga:=#1\do
                   {\expandafter\@setparms\@psfiga,}%
\if@prologfile
\fi
}%
\def\bb@missing{%
        \if@verbose
            \typeout{psfig: searching \@p@sfile \space  for bounding box}%
        \fi
        \epsfgetbb{\@p@sfile}%
        \ifepsfbbfound
            \ps@dimcent=\epsfllx bp\edef\@p@sbbllx{\number\ps@dimcent}%
            \ps@dimcent=\epsflly bp\edef\@p@sbblly{\number\ps@dimcent}%
            \ps@dimcent=\epsfurx bp\edef\@p@sbburx{\number\ps@dimcent}%
            \ps@dimcent=\epsfury bp\edef\@p@sbbury{\number\ps@dimcent}%
        \else
            \epsfbbfoundfalse
        \fi
}
\newdimen\p@intvaluex
\newdimen\p@intvaluey
\def\rotate@#1#2{{\dimen0=#1 sp\dimen1=#2 sp
                  \global\p@intvaluex=\cosine\dimen0
                  \dimen3=\sine\dimen1
                  \global\advance\p@intvaluex by -\dimen3
                  \global\p@intvaluey=\sine\dimen0
                  \dimen3=\cosine\dimen1
                  \global\advance\p@intvaluey by \dimen3
                  }}%
\def\compute@bb{%
                \epsfbbfoundfalse
                \if@bbllx\epsfbbfoundtrue\fi
                \if@bblly\epsfbbfoundtrue\fi
                \if@bburx\epsfbbfoundtrue\fi
                \if@bbury\epsfbbfoundtrue\fi
                \ifepsfbbfound\else\bb@missing\fi
                \ifepsfbbfound\else
                \@latexerr{ERROR. cannot locate BoundingBox}\@whattodobb
                        \@p@@sbbllx{100bp}%
                        \@p@@sbblly{100bp}%
                        \@p@@sbburx{200bp}%
                        \@p@@sbbury{200bp}%
                        \no@bbtrue
                        \psdraft
                \fi
                \count203=\@p@sbburx
                \count204=\@p@sbbury
                \advance\count203 by -\@p@sbbllx
                \advance\count204 by -\@p@sbblly
                \edef\ps@bbw{\number\count203}%
                \edef\ps@bbh{\number\count204}%
                 \edef\@bbw{\number\count203}%
                \edef\@bbh{\number\count204}%
               \if@angle
                        \Sine{\@p@sangle}\Cosine{\@p@sangle}%
 
{\ps@dimcent=\maxdimen\xdef\r@p@sbbllx{\number\ps@dimcent}%
 
\xdef\r@p@sbblly{\number\ps@dimcent}%
 
\xdef\r@p@sbburx{-\number\ps@dimcent}%
 
\xdef\r@p@sbbury{-\number\ps@dimcent}}%
                        \def\minmaxtest{%
                           \ifnum\number\p@intvaluex<\r@p@sbbllx
                              \xdef\r@p@sbbllx{\number\p@intvaluex}\fi
                           \ifnum\number\p@intvaluex>\r@p@sbburx
                              \xdef\r@p@sbburx{\number\p@intvaluex}\fi
                           \ifnum\number\p@intvaluey<\r@p@sbblly
                              \xdef\r@p@sbblly{\number\p@intvaluey}\fi
                           \ifnum\number\p@intvaluey>\r@p@sbbury
                              \xdef\r@p@sbbury{\number\p@intvaluey}\fi
                           }%
                        \rotate@{\@p@sbbllx}{\@p@sbblly}%
                        \minmaxtest
                        \rotate@{\@p@sbbllx}{\@p@sbbury}%
                        \minmaxtest
                        \rotate@{\@p@sbburx}{\@p@sbblly}%
                        \minmaxtest
                        \rotate@{\@p@sbburx}{\@p@sbbury}%
                        \minmaxtest
 
\edef\@p@sbbllx{\r@p@sbbllx}\edef\@p@sbblly{\r@p@sbblly}%
 
\edef\@p@sbburx{\r@p@sbburx}\edef\@p@sbbury{\r@p@sbbury}%
                \fi
                \count203=\@p@sbburx
                \count204=\@p@sbbury
                \advance\count203 by -\@p@sbbllx
                \advance\count204 by -\@p@sbblly
                \edef\@bbw{\number\count203}%
                \edef\@bbh{\number\count204}%
}%
\def\in@hundreds#1#2#3{\count240=#2 \count241=#3
                     \count100=\count240        
                     \divide\count100 by \count241
                     \count101=\count100
                     \multiply\count101 by \count241
                     \advance\count240 by -\count101
                     \multiply\count240 by 10
                     \count101=\count240        
                     \divide\count101 by \count241
                     \count102=\count101
                     \multiply\count102 by \count241
                     \advance\count240 by -\count102
                     \multiply\count240 by 10
                     \count102=\count240        
                     \divide\count102 by \count241
                     \count200=#1\count205=0
                     \count201=\count200
                        \multiply\count201 by \count100
                        \advance\count205 by \count201
                     \count201=\count200
                        \divide\count201 by 10
                        \multiply\count201 by \count101
                        \advance\count205 by \count201
                     \count201=\count200
                        \divide\count201 by 100
                        \multiply\count201 by \count102
                        \advance\count205 by \count201
                     \edef\@result{\number\count205}%
}%
\def\compute@wfromh{%
                \in@hundreds{\@p@sheight}{\@bbw}{\@bbh}%
                \edef\@p@swidth{\@result}%
}%
\def\compute@hfromw{%
                \in@hundreds{\@p@swidth}{\@bbh}{\@bbw}%
                \edef\@p@sheight{\@result}%
}%
\def\compute@handw{%
                \if@height
                        \if@width
                        \else
                                \compute@wfromh
                        \fi
                \else
                        \if@width
                                \compute@hfromw
                        \else
                                \edef\@p@sheight{\@bbh}%
                                \edef\@p@swidth{\@bbw}%
                        \fi
                \fi
}%
\def\compute@resv{%
                \if@rheight \else \edef\@p@srheight{\@p@sheight} \fi
                \if@rwidth \else \edef\@p@srwidth{\@p@swidth} \fi
}%
\def\compute@sizes{%
        \if@scalefirst\if@angle
        \if@width
           \in@hundreds{\@p@swidth}{\@bbw}{\ps@bbw}%
           \edef\@p@swidth{\@result}%
        \fi
        \if@height
           \in@hundreds{\@p@sheight}{\@bbh}{\ps@bbh}%
           \edef\@p@sheight{\@result}%
        \fi
        \fi\fi
        \compute@handw
        \compute@resv
}
\long\def\graphic@verb#1{\def\next{#1}%
  {\expandafter\graphic@strip\meaning\next}}
\def\graphic@strip#1>{}
\def\graphic@zapspace#1{%
  #1\ifx\graphic@zapspace#1\graphic@zapspace%
  \else\expandafter\graphic@zapspace%
  \fi}
\def\psfig#1{%
\edef\@tempa{\graphic@zapspace#1{}}%
\ifvmode\leavevmode\fi\vbox {%
        \ps@init@parms
        \parse@ps@parms{\@tempa}%
        \ifnum\@psdraft=1
                \typeout{[\@p@sfilefinal]}%
                \if@verbose
                        \typeout{epsfig: using PSFIG macros}%
                \fi
                \psfig@method
        \else
                \epsfig@draft
        \fi
}
}%
\def\graphic@zapspace#1{%
  #1\ifx\graphic@zapspace#1\graphic@zapspace%
  \else\expandafter\graphic@zapspace%
  \fi}
\def\epsfig#1{%
\edef\@tempa{\graphic@zapspace#1{}}%
\ifvmode\leavevmode\fi\vbox {%
        \ps@init@parms
        \parse@ps@parms{\@tempa}%
        \ifnum\@psdraft=1
          \if@angle\use@psfigtrue\fi
          {\ifnum\fig@driver=1\global\use@psfigtrue\fi}%
          {\ifnum\fig@driver=3\global\use@psfigtrue\fi}%
          {\ifnum\fig@driver=4\global\use@psfigtrue\fi}%
          {\ifnum\fig@driver=5\global\use@psfigtrue\fi}%
                \ifuse@psfig
                        \if@verbose
                                \typeout{epsfig: using PSFIG macros}%
                        \fi
                        \psfig@method
                \else
                        \if@verbose
                                \typeout{epsfig: using EPSF macros}%
                        \fi
                        \epsf@method
                \fi
        \else
                \epsfig@draft
        \fi
}%
}%

\def\epsf@method{%
        \epsfbbfoundfalse
        \if@bbllx\epsfbbfoundtrue\fi
        \if@bblly\epsfbbfoundtrue\fi
        \if@bburx\epsfbbfoundtrue\fi
        \if@bbury\epsfbbfoundtrue\fi
        \ifepsfbbfound\else\epsfgetbb{\@p@sfile}\fi
        \ifepsfbbfound
           \typeout{<\@p@sfilefinal>}%
           \epsfig@gofer
        \else
          \@latexerr{ERROR - Cannot locate BoundingBox}\@whattodobb
          \@p@@sbbllx{100bp}%
          \@p@@sbblly{100bp}%
          \@p@@sbburx{200bp}%
          \@p@@sbbury{200bp}%
                \count203=\@p@sbburx
                \count204=\@p@sbbury
                \advance\count203 by -\@p@sbbllx
                \advance\count204 by -\@p@sbblly
                \edef\@bbw{\number\count203}%
                \edef\@bbh{\number\count204}%
          \compute@sizes
          \epsfig@@draft
       \fi
}%
\def\psfig@method{%
        \compute@bb
        \ifepsfbbfound
          \compute@sizes
          \psfig@start
          \vbox to \@p@srheight sp{\hbox to \@p@srwidth 
            sp{\hss}\vss\psfig@end}%
        \else
           \epsfig@draft
        \fi
}%
\def\epsfig@draft{\compute@bb\compute@sizes\epsfig@@draft}%
\def\epsfig@@draft{%
\typeout{<(draft only) \@p@sfilefinal>}%
\if@draftbox
        \hbox{{\fboxsep0pt\fbox{\vbox to \@p@srheight sp{%
        \vss\hbox to \@p@srwidth sp{ \hss 
           \expandafter\Literally\@p@sfilefinal\@nil
                          \hss }\vss
        }}}}%
\else
        \vbox to \@p@srheight sp{%
        \vss\hbox to \@p@srwidth sp{\hss}\vss}%
\fi
}%
\def\Literally#1\@nil{{\tt\graphic@verb{#1}}}
\psfigdriver{dvips}%
\epsfigRestoreAt

\def\newpic#1{%
   \def\emline##1##2##3##4##5##6{%
      \put(##1,##2){\special{em:point #1##3}}%
      \put(##4,##5){\special{em:point #1##6}}%
      \special{em:line #1##3,#1##6}}}
\newpic{}

\def\be{\begin{equation}}
\def\ee{\end{equation}}
\def\bea{\begin{eqnarray}}
\def\eea{\end{eqnarray}}

              \def\d{\delta}
    \def\f{\phi}       
   \def\G{\Gamma}        \def\l{\lambda}
  \def\m{\mu}              
      \def\p{\psi}       \def\P{\Psi}
       \def\t{\tau}
\def\vf{\varphi}             

\def\CD{{\cal D}}

\def\ha{{1\over2}}
\newcommand{\ba}[1]{\left(\begin{array}{#1}}
\newcommand{\ea}{\end{array}\right)}
\newcommand{\fr}[2]{\frac{#1}{#2}}
\newcommand{\nn}{\nonumber}

\def\tr{\hbox{Tr}\,}
\def\tildef{\tilde{\varphi}}

\def\dag{\dagger}

\def\prd#1#2#3{Phys. Rev. D {\bf {#1}}, {#2} ({#3})}
\def\prs#1#2#3{Phys. Rev. {\bf {#1}}, {#2} ({#3})}

\def\plb#1#2#3{Phys. Lett. {\bf B{#1}}, {#2} ({#3})}
\def\pla#1#2#3{Phys. Lett. {\bf A{#1}}, {#2} ({#3})}
\def\npb#1#2#3{Nucl. Phys. {\bf B{#1}}, {#2} ({#3})}

\def\ap#1#2#3{Ann. Phys. {\bf {#1}}, {#2} ({#3})}

\def\zp#1#2#3{Z. Phys. {\bf {#1}}, {#2} ({#3})}

\begin{document}

\title{Variational Approach to Quantum Field Theory: 
Gaussian Approximation and the Perturbative Expansion around It\footnote{To appear
in the proceedings of APCTP-ICTP Joint International Conference '97 on
{\it Recent Developments in Nonperturbative Quantum Field Theory}, May 26-30, 1997, Seoul}}

\author{Jae Hyung Yee}

\address{Department of Physics and Institute for Mathematical Sciences \\
Yonsei University, Seoul 120-749, Korea \\E-mail: jhyee@phya.yonsei.ac.kr}


\maketitle\abstracts{
The functional Schr\"odinger picture formulation of quantum field theory and the variational 
Gaussian approximation method based on the formulation are briefly reviewed.
After presenting recent attempts to improve the variational approximation, we introduce
a new systematic method based on the background field method, 
which enables one to compute the order-by-order correction terms 
to the Gaussian approximation of the effective action.
}

\newpage

\section{Introduction} 

The variational approximation method has been proved to be a convenient
 non-perturbative approximation method for both the quantum mechanical
 and quantum field theoretic systems. The method, which was originally
 developed for a non-perturbative evaluation of quantum mechanical energy eigenvalues, has been extended to an action principle 
by Jackiw and Kerman\cite{r1}, 
which enables one to compute the effective actions for time-dependent 
systems and also for quantum field theories. 
The usefulness of the method for quantum field theories, however, 
is limited by the fact that only the Gaussian functional and it's obvious generalizations can be used as trial wave functionals, which is the reason 
why it is often called the Gaussian approximation.
Attempts to improve the Gaussian approximation by using more general non-Gaussian trial wave functionals have not been as fruitful as expected,
due to the difficulties in solving the variational gap equations\cite{r2}.
To overcome such limitations the authors of Ref.[3] has developed 
the so-called post-Gaussian expansion method,
which enables one to compute the order-by-order correction terms to the Gaussian effective potentials for the time-independent systems.

It is the purpose of this note to give a brief review of the variational approximation method in quantum field theory, and to present a new systematic method of perturbative expansion around the Gaussian effective action based on 
the background field method, which is also applicable to the time-dependent systems.

The variational method in quantum mechanics is based on the fact 
that the ground state energy is the minimum of the expectation values of
the Hamiltonian of the system. This can be written as a variational principle,
\be 
\d_\P <\P|\hat{H}(\hat{Q},\hat{P}) - \l |\P> = 0,
\label{aaa}
\ee
where $\hat{H}$ is the Hamiltonian of the system and $|\P>$ is an arbitrary state.
If one starts from a trial wave function with variational parameters, 
then Eq.(\ref{aaa}) provides a variational approximation of the energy eigenstates.

This simple variational method can be generalized to include the information 
on the time-evolution of the system, by choosing the state to satisfy the constraints,
\bea
<\P|\P> &=& 1  \nn  \\
<\P|\hat{Q}|\P> &=& q(t)
{\label{aab}} \\
<\P|\hat{P}|\P> &=& p(t). \nn 
\eea
Minimizing the expectation value of the Hamiltonian with respect to 
such states,
\be
\d_\P<\P|\hat{H}(\hat{Q},\hat{P})|\P> = 0,
\label{aac}
\ee
we obtain the effective Hamiltonian,
\be
H_{eff} (q,p) = <\P_0|\hat{H}(\hat{Q},\hat{P})|\P_0>,
\label{aad}
\ee
where $|\P_0>$ is the solution of Eq.(\ref{aac}). 
This effective Hamiltonian provides many useful informations on the quantum system. 
If one minimizes $H_{eff}$ with respect to the variations of $q(t)$ and $p(t)$, 
one obtains the ground state energy eigenvalue of the system. 
If one solves the Hamilton's equation, on the other hand,
\bea
 \dot{q}(t) &=& \fr{\partial H_{eff}}{\partial p}  \label{aae}
  \\
 \dot{p}(t) &=& -\fr{\partial H_{eff}}{\partial q},  \nn
\eea
the solution $q$ and $p$ give the quantum mechanical informations on
$<\P_0|\hat{Q}|\P_0>$ and $<\P_0|\hat{P}|\P_0>$, respectively.

If we set $\dot{q} = 0 = p$ in Eq.(\ref{aab}), we obtain the effective potential,
\be
V_{eff}(q) = H_{eff}(q,p)|_{p=0}.
\label{aaf}
\ee

If one starts from a trial wave function which satisfies Eq.(\ref{aab}), 
then one obtains the variational approximation of the effective Hamiltonian 
or the effective potential.

This simple argument is in fact a part of the quantum action principle 
in the Schr\"{o}dinger picture called the action principle of Dirac\cite{r1,r4}.
This formulation can be generalized to the case of quantum field theory, 
which is the functional Schr\"{o}dinger picture formulation of quantum field theory\cite{r4}. 

\section{Functional Schr\"{o}dinger picture representation of 
quantum field theory and the variational method.}

\subsection{Schr\"{o}dinger picture representation of scalar field theory}

The scalar field theory is defined by the Lagrangian density,
\be
{\cal L} = \fr{1}{2}(\pi\dot{\phi} - \phi\dot{\pi}) - {\cal H}[\phi,\pi],
\label{baa}
\ee
and the equal-time commutation relations
\bea
 [ \phi(x) , \pi(x') ]_{x^0 = x'^0}  &=& i \d ( \vec{x} - \vec{x}' ) \label{bab} \\\
 [ \phi(x) , \phi(x') ]_{x^0 = x'^0}  &=&  0 = [\pi(x), \pi(x')]_{x^0 = x'^0}, \nn
\eea
where ${\cal H}[\phi,\pi] = \fr{1}{2} \pi^2 + V(\phi)$ is the Hamiltonian density.

As in the quantum mechanical case, we choose the basis vectors
for the Schr\"{o}dinger representation to be the eigenstates of the field operator $\phi(x)$,
\be
\phi | \varphi(\vec{x})> = \varphi(\vec{x}) | \varphi(\vec{x})>,
\label{bac}
\ee
which satisfy the orthogonality relation and the completeness relation,
\bea
<\varphi | \varphi '> = \d[\varphi (\vec{x}) - \varphi '(\vec{x}') ] \label{bad} \\
\int {\cal D}\varphi |\varphi><\varphi| = I. \nn
\eea

Then the state of the physical system is represented by the component,
$\Psi[\varphi, t] = \;\;<\varphi | \Psi>$, 
of the expansion,
\be
|\Psi> = \int {\cal D}\varphi |\varphi><\varphi|\Psi>,
\label{bae}
\ee
and the operators are represented by the matrix elements,
\bea
<\varphi|\phi|\varphi'> &=& \varphi(\vec{x}) \delta[\varphi - \varphi'] \label{baf} \\
<\varphi|\pi|\varphi'> &=& -i \fr{\delta}{\delta \varphi(\vec{x})} \delta
[\varphi - \varphi'], \nn
\eea
which are obtained from the matrix elements of Eq.(\ref{bab}).
The time-evolution of the system is then determined by the functional 
Schr\"{o}dinger equation,
\bea
i \fr{d}{dt}\Psi[\varphi, t] &=& <\varphi|H[\phi,\pi]|\varphi'><\varphi'|\Psi> 
\nn \\
&=& H[\varphi, -i \fr{\delta}{\delta\varphi}] \Psi[\varphi, t].
\label{bag}
\eea
This is a simple generalization of the Schr\"{o}dinger picture representation
of quantum mechanics to the case of the systems with infinite degrees of freedom.

\subsection{Free Scalar field theory}

To obtain physical information in the Schr\"{o}dinger picture formulation one has to 
solve the functional Schr\"{o}dinger equation (\ref{bag}). As an example of 
solving the equation of motion, we consider the free scalar field theory 
described by the Hamiltonian density,
\be
{\cal H} = \fr{1}{2} \pi^2 + \fr{1}{2}(\nabla\phi)^2 + \fr{1}{2} m^2\phi^2.
\label{bah}
\ee
By using the representation (\ref{baf}), the Hamiltonian of the system can be written as 
\be
H = \fr{1}{2} \int d^3 x [- \fr{\delta}{\delta\varphi(\vec{x}) }\fr{\delta}
{\delta\varphi(\vec{x}) }- \varphi(\vec{x}) (\nabla^2 - m^2 )\varphi(\vec{x}) ].
\label{bai}
\ee

Since the Hamiltonian has no explicit time-dependence, there exist stationary
states satisfying the energy eigenvalue equation,
\be
H[\varphi, -i\fr{\delta}{\delta\varphi}]\Psi[\varphi] = E \Psi[\varphi],
\label{baj}
\ee
where $E$ is the energy eigenvalue. Since the Hamiltonian (\ref{bai}) is 
quadratic in 
$\varphi(\vec{x})$,
the Gaussian functional 
\be
\Psi[\varphi] = e^{-\fr{1}{2} \int d^3 xd^3 x' \varphi(\vec{x}) G(\vec{x},
\vec{x}' ) \varphi(\vec{x} ')},
\label{bak}
\ee
is a solution of (\ref{baj}). Substituting (\ref{bak}) into (\ref{baj}),
we obtain.
\be
\fr{1}{2} [G(\vec{x},\vec{x}') - \varphi(\vec{x})G^2 (\vec{x},\vec{x}')
\varphi(\vec{x}') + \varphi(\vec{x})(-\nabla^2 + m^2 )\varphi(\vec{x}) ]
\Psi[\varphi] = E\Psi[\varphi],
\label{bal}
\ee
where the repeated arguments imply the integration over the variables. 
For this equation to be true for all eigenvalue $\varphi(\vec{x})$,
the following relations must be satisfied,
\bea
E  =  \fr {1}{2} Tr G &=& \fr {1}{2} \int d^3 x G(\vec{x}, \vec{x} ) \label{bam}  \\
G^2 ( \vec{x},\vec{x}' )  &=& (- \nabla^2 + m^2 ) \delta (\vec{x} - \vec{x}'). \nn
\eea
Solving the second equation of (\ref{bam}) via Fourier transformation,
one finds the kernel of the Gaussian functional (\ref{bak}) is given by
\be
G^{-1} (\vec{x},\vec{x}') = \fr{1}{(2\pi)^3 } \int d^3 p e^{i \vec{p} \cdot
(\vec{x} - \vec{x}') } \fr{1}{\sqrt{ \vec{p}^2 + m^2 }},
\label{ban}
\ee
which is the propagator of the free scalar field theory at equal-time.
From this the eigenvalue $E$ is found to be 
\be
E = \fr{1}{2} \fr{1}{(2\pi)^3 } \int d^3 x \int d^3 p \sqrt{\vec{p}^2 + m^2 },
\label{bao}
\ee
which is the well-known zero-point energy of the free scalar field theory.

We thus find that the ground state of the free scalar field theory 
is described by the Gaussian wave functional with the kernel given by the 
inverse propagator at equal-time. 
One can construct all the eigenstates of the Hamiltonian by considering 
the creation and annihilation operators.
They are given by the Hermite polynomial functionals multiplied by the 
ground(vacuum) state\cite{r5}.

For the interacting theory, however, we cannot solve the functional 
Schr\"{o}dinger equation (\ref{bag}) in closed form. 
One thus has to use 
an approximation method to find solutions to the equation of motion.
A natural way to evaluate an approximate solution to (\ref{bag}) in the Schr\"{o}dinger 
picture is to use the variational method, which is the subject of the next
subsection.

\subsection{Gaussian approximation of $\phi^4$ theory}

 To illustrate the variational approximation method in quantum field theory, 
we now consider the scalar $\phi^4$ theory described by the Hamiltonian 
density,
\be 
{\cal H} =\fr{1}{2}\pi^2 + \fr{1}{2}(\vec{\nabla}\phi)^2 + 
\fr{1}{2}\mu^2\phi^2 + \fr{\lambda}{4!}\phi^4  , 
\ee
 and compute the Gaussian approximation of the effective potential\cite{r6}. 
To do this we take as a trial wave functional the solution of a free field 
theory discussed in the last subsection, namely, the Gaussian wave functional,
\be
 \P[\vf,t] = Ne^{-\fr{1}{4}[\vf(\vec{x})-
\hat{\phi}]G^{-1}(\vec{x},\vec{x}';t)
[\vf(\vec{x}')-\hat{\phi}]+\fr{i}{\hbar}\hat{\pi}(\vec{x},t)[\vf(\vec{x})-\hat{\phi}]}
\label{2.17}
\ee
 which satisfies the constraints,
\bea
 <\Psi\mid \Psi> &=& 1 \nonumber \\
 <\Psi\mid \phi \mid \Psi> &=& \hat{\phi}(\vec{x},t) \label{2.18} \\
 <\Psi\mid \pi \mid \Psi> &=& \hat{\pi}(\vec{x},t). \nn
\eea

 Since the effective potential is the effective Hamiltonian with vanishing 
momenta, we set $\hat{\pi}$ equal to zero and consider the space-time 
independent $\hat{\phi}$. Then the expectation value of the Hamiltonian with
respect to the trial state(\ref{2.17}) is given by
\bea
 <H> &=&  \int  d^3 x \left(  V(\hat{\phi}) + \hbar[\fr{1}{8}G^{-1}(\vec{x},\vec{x})
-\fr{1}{2}\nabla^2G(\vec{x},\vec{x}')|_{\vec{x}=\vec{x}'} \right.  \nn \\ 
&& +  \left. V^{(2)}(\hat\phi)G(\vec{x},\vec{x})]
+ \fr{\hbar^2}{8}V^{(4)}(\hat\phi) ( G(\vec{x},\vec{x}) )^2  \right),
\label{2.19}     
\eea
where $V(\hat\phi)=\fr{\mu^2}{2}{\hat\phi}^2+\fr{\lambda}{4!}{\hat\phi}^4$ is 
the classical potential and $V^{(n)}(\hat\phi)= (\fr{d}{d\hat\phi})^nV(\hat\phi)$.

The variational equation,
\be
\delta_G<H> =0,
\ee
yields the variational gap equation,
\be
\fr{1}{4}G^{-2}(\vec{x},\vec{x}') =[ -\nabla^2 + V^{(2)}(\hat\phi)+\fr{\hbar}{2}V^{(4)}
(\hat\phi)G(\vec{x},\vec{x})]\delta(\vec{x}-\vec{x}'). 
\label{2.21}
\ee 
Substituting(\ref{2.21}) into (\ref{2.19}) one obtains the effective potential,
\be
V_{eff}(\hat\phi) = \fr{1}{2}\mu^2{\hat\phi}^2+\fr{\lambda}{4!}{\hat\phi}^4 + \fr{\hbar}{4}
G^{-1}(\vec{x},\vec{x})-\fr{\hbar^2}{8}G(\vec{x},\vec{x})G(\vec{x},\vec{x}).
\label{2.22}
\ee
The first three terms Eq.(\ref{2.22}) constitute the one-loop effective potential, and
the last term contains the higher-loop contribution, which shows the non-perturbative
nature of the approximation.

Using the Fourier transformation method we can solve the gap equation(\ref{2.21}):
\be
 G(\vec{x}-{\vec{x}}')=\int{\fr{d^3p}{{(2\pi)}^3}e^{i\vec{p} \cdot (\vec{x}-\vec{x}')}
\fr{1}{\sqrt{ {\vec{p}}^2+m^2 } } },
\label{gggg1}
\ee
where the effective mass m is defined by
\be
m^2 = \mu^2 + \fr{\lambda}{2}{\hat\phi}^2 + \fr{\lambda}{2}\hbar\int \fr{d^3p}{ {(2\pi)}^3 } 
\fr{1}{\sqrt{ {\vec{p}}^2+m^2 }}.
\label{2.24}
\ee
Substituting these results into (\ref{2.22}) yields the effective potential:
\bea                                                  
V_{eff}(\hat\phi)&=&\fr{\mu^2}{4}{\hat\phi}^2 + \fr{\lambda}{4!}{\hat\phi}^4 
- \fr{1}{2\lambda}{\left(m^2-\mu^2-\fr{\lambda}{2}{\hat\phi}^2 \right )}^2 + \fr{m^2}{2}I_1 \nn \\
&-&\fr{m^4}{4}I_2(M)+\fr{m^4}{64\pi^2}\left(\ln{\fr{m^2}{M^2}}-\fr{1}{2} \right),
\label{2.25}
\eea
where  
\be
m^2 = \mu^2+\fr{\lambda}{2}I_1+\fr{\lambda}{2}{\hat\phi}^2
-\fr{\lambda}{2}m^2I_2(M)+\fr{\lambda m^2}{32\pi^2} \ln \fr{m^2}{M^2},
\ee
and, $I_1$ and $I_2 (M)$ are given by 
\be
I_1 = \int{\fr{d^3p}{ {(2\pi)}^3 }\fr{1}{2|\vec{p}|}} \;\;\; , \;\;\;\;
I_2(M)=\int{\fr{d^3p}{ {(2\pi)}^3 }\left(\fr{1}{2|\vec{p}|} 
-\fr{1}{2\sqrt{{\vec{p}}^2+\mu^2} } \right) }.
\ee

 The effective potential (\ref{2.25}) has two divergences, $I_1$ and $I_2(M)$,
and thus we need to renormalize the theory. The renormalization conditions 
can be written as [6],
\be
\left. \fr{dV_{eff}}{dm^2} \right|_{m_0} = \fr{\mu_R^2}{\lambda_R} \; \; \;, \; \; \; \; 
\left. \fr{d^2V_{eff}}{d{(m^2)}^2} \right|_{m_0} = \fr{1}{3\lambda_{R}},
\label{2.26}
\ee
which, in our case, yields
\bea
\fr{\mu_R^2}{\lambda_R} &=& \fr{\mu^2}{\lambda} + \fr{1}{2}I_1  \label{2.27} \\
\fr{1}{\lambda_R} &=& \fr{1}{\lambda} + \fr{1}{2}I_2(M). \nn
\eea
 Substituting (\ref{2.27}) into (\ref{2.25}), we finally obtain the renormalized
 effective potential,
\be
 V_{eff}(\hat\phi) = -\fr{m^4}{2\lambda_R} + \fr{m^2}{2}{\hat\phi}^2 +
\fr{\mu_R^2}{\lambda_R}m^2 + \fr{m^4}{64\pi^2}
\left(\ln\fr{m^2}{M^2} - \fr{1}{2} \right),
\label{2.28}
\ee
where
\be
m^2 = \mu_R^2 + \fr{\lambda_R}{2}{\hat\phi}^2 + \fr{\lambda_R}{32\pi^2} 
\ln\fr{m^2}{M^2} .
\label{2.29}
\ee

 This result is qualitatively equivalent to the large-$N$ approximation 
(where $N$ is the number of the scalar field components)\cite{r7}, and shows the
 non-perturbative nature of the approximation. 
The result (\ref{2.28})
 can be used to study the quantum nature of the theory, such as the
symmetry breaking phenomena\cite{r6}. The method can be generalized to the case
of the field theories in curved space\cite{r8}, to the finite temperature
quantum field theory\cite{r9}, and also to the systems out of thermal 
equilibrium\cite{r10}.

\subsection{Fermionic theory}

 Finding the Schr\"{o}dinger picture representation of fermion fields is non-trivial
since the equal-time anti-commutation relation, in Hermitian representation,
\be
\{\psi_{\alpha }(x), \psi_{\beta}(x') \}_{x^0={x^0}' } = \delta_{\alpha\beta}\delta
(\vec{x}-\vec{x}'),
\ee
implies that the field operators $\psi$'s are their own momentum conjugates.
In the non-Hermitian representation, on the other hand, the anti-commutation
relation is given by 
\be
\{\psi_{\alpha}(x), \psi_{\beta}^\dag(x')\}_{x^0={x^0}' } = \delta_{\alpha\beta}\delta
(\vec{x}-\vec{x}'),
\label{2.31}
\ee
but now the eigenstates of field operators do not form an orthonormal basis.
One can show, however, that there exists a one parameter family of equivalent 
representations of fermion fields\cite{r4}.
Floreanini and Jackiw\cite{r11}, in particular, have found that the representation,
\bea
\psi=\fr{1}{\sqrt2}[\theta(\vec{x}) + \fr{\delta}{\delta\theta^\dag(\vec{x})}] 
\label{2.32} \\       
\psi^\dag=\fr{1}{\sqrt2}[\theta^\dag(\vec{x}) + \fr{\delta}{\delta\theta(\vec{x})}],
\nn
\eea
where $\theta(\vec{x})$ is a complex Grassmann variable,
satisfies the anti-commutation relation (\ref{2.31}).

Using this representation one can follow the same procedure as in the case
of the  scalar field theory to evaluate the Gaussian  approximation of the
Fermionic field theories.

This method has been applied to the cases of (1+1)-dimensional and (2+1)-
dimensional Gross-Neveu models\cite{r12}, and found that the results are 
qualitatively equivalent to those of the large-$N$ approximation. It has
also been applied to the (2+1)-dimensional Thirring model, and in this case
the Gaussian result has been found to be qualitatively better than the large-$N$
 approximation\cite{r13}.
The Gaussian effective potential shows the existence of the symmetry breaking
in the theory, while the large-$N$ does not. The method has also been found to 
be effective in determining the existence of the bound states in the theory\cite{r14}.

\section{Beyond Gaussian approximation?}

 We have shown that the Gaussian approximation based on the functional 
Schr\"{o}dinger picture formulation is a simple generalization of the quantum
mechanical variational method, and provides a convenient non-perturbative 
approximation method for various quantum field theories including the systems
out of thermal equilibrium. However, the method has a serious drawback in that,
unlike the quantum mechanical cases, one can compute the expectation value of
the quantum field theoretic Hamiltonian only for the Gaussian wave functional
and  its simple generalizations.

If one could develop a systematic method to compute the correction terms to 
this approximation, therefore, the variational method would become a very 
powerful technique in obtaining physical information from quantum field
theories. In fact there have been some attempts in this direction recently.

One of such attempts, known as the post-Gaussian approximation method or the variational
perturbation theory\cite{r3}, is based on the observation that the Gaussian wave 
functional
\be
\Psi_0 = e^{-\fr{1}{4}\tilde{\varphi}(\vec{x})G^{-1}(\vec{x}-\vec{x}')\tilde{\varphi}
(\vec{x}')}\;\;\;,\;\;\;\; \; \tilde{\varphi} = \varphi-\hat\phi
\label{3.1}
\ee
is a vacuum state for a free theory with mass given by
\be
m^2 = \mu^2 + \fr{\lambda}{2}{\hat\phi}^2 + \fr{\lambda}{2}\int
\fr{d^3p}{{(2\pi)}^3}g(p),
\label{3.2}
\ee
where $g(p)$ is determined by the variational gap equation (\ref{gggg1}) and (\ref{2.24}). 

One can construct
all the eigenstates of this free theory:
\be
\Psi_n[\tilde{\varphi}] = H_n[\tilde{\varphi}]\Psi_0[\tilde{\varphi}]\;\;\;, \;\;\;\;\;
n= 0, 1, 2,\cdots
\label{3.3}
\ee
where $H_n[\tilde{\varphi}]$ is the Hermite polynomial functional.
One splits the Hamiltonian as
\be
H=H_0 + H_I
\label{3.4}
\ee
where $H_0$ is the free Hamiltonian with the mass $m$ given by Eq.(\ref{3.2})
and $H_I$ is the remaining part of $H$, and then applies the well known
perturbation theory with the states (\ref{3.3}) taken to be the basis for
the perturbation.

 The effective potential is the expectation value of the Hamiltonian with
the constraint
\be
<\Psi_0|\phi|\Psi_0> = \hat\phi,
\label{3.5}
\ee
where $|\Psi_0>$ is the ground state of the system.

Taking the eigenstates (\ref{3.3}) to be the basis for perturbative expansion,
the effective potential can be written as
\be
V_{eff}(\hat\phi) = E_0(\hat\phi) + \sum^\infty_{n=0}<\Psi_0|\Psi_I
 \left( \fr{1}{E_0-H_0}H_I \right)^n|\Psi_0>,
\label{3.6}
\ee
where the first term is the Gaussian effective potential and the second term
represents the perturbative correction terms\cite{r3}. This method has been used to
compute the first non-trivial contribution to the Gaussian effective potential
for the (0+1)-dimensional $\phi^4$ theory by Jeon\cite{r15} and the (1+1)- and
(2+1)-dimensional $\phi^4$ theories by Cea and Tedesco\cite{r3}.

 This method provides a conceptually simple procedure to compute the order-by-order
correction terms to the Gaussian effective potential for time-independent
systems. For the time-dependent systems such as those out of thermal 
equilibrium\cite{r10}, however, we need a method that can deal with the effective 
action, which contains information on the time-evolution of the system.
It is the subject of the next section to introduce a new systematic expansion
method for effective action, which was developed by Geon Hyoung Lee\cite{r16}.

\section{The Background field method}   
We now introduce a systematic method which enables one to compute 
the order by order correction terms to the Gaussian approximation
of the effective action, that contains the information on the 
time-evolution of the system.   
This new method is based on the background
field method. 
So we will start to give a brief review on the background 
field method\cite{r17}.
\subsection{Background field method}
The generating functional for the connected Green's functions, $W[J]$, is 
defined by the path integral,
\be
 e^{iW[J]}=N\int {\cal D}\f e^{iS[\f]+iJ\f},
\ee   
where $S[\f]$ is the classical action, and the integral convention,
 $J\f\equiv \int d^nxJ(x)\f(x)$ in $n$-dimensional space-time, is used in the exponent. 
The effective action
$\Gamma[\f]$ is then defined by the Legendre transformation, 
\be
\Gamma[\vf]=W[J]-J\vf,
\ee
where $\vf(x)$ is the vacuum expectation value of the field operator,
\be
\vf(x)=<\f(x)>_J=\fr {\d}{\d J(x)} W[J].
\ee

The background field method is one of the most effective ways of computing the
effective action in the loop expansion, and is obtained by introducing the 
background field  $B(x)$ to the classical action:
\be
S[\f] \rightarrow S[\f+B].
\ee
This defines a new generating
functional,
\bea
e^{i\tilde{W}[J,B]}&=&N\int {\cal D}\f e^{iS[\f+B]+iJ\f}  \nn   \\
&=& N\int{\cal D}\f' e^{iS[\f']+iJ[\f'-B]}.
\label{test1}
\eea 
Eg.(\ref{test1}) implies that this new generating functional is related to the 
old one by 
\be
\tilde{W}[J,B]=W[J]-JB,
\label{test2}
\ee
and the vacuum expectation value of the field operator in the presence of the 
background field is given by,
\be
\tilde{\vf}=\fr {\delta \tilde{W}} {\delta J}=\vf-B.
\ee 

Now consider the case when the vacuum expectation value of the field operator 
in the presense of 
the background field, $\tilde{\vf}$, vanishes:
\be
\fr {\delta \tilde{W}} {\delta J}=\tilde{\vf}=0.
\ee 
This determines the configuration of $J$ which extremizes the generating 
functional $\tilde{W}[J,B]$:
\be
 J=J[B].
\ee
Then one can easily show from Eq.(\ref{test2}) that the extremum value of
the generating functional $\tilde{W}$ becomes the effective action:
\be
\tilde{W}[J,B]|_{\tilde{\vf}=0}=W[J(B)]-J(B)B=\Gamma[B].
\ee

The simpler way to compute the effective action $\Gamma[B]$ is to consider 
the Legendre transformation of $\tilde{W}[J,B]$:
\be
\tilde{\Gamma}[\tilde{\vf},B]=\tilde{W}[J,B]-J\tilde{\vf},
\label{test3}
\ee
which generates the one $\tilde{\vf}$-particle irreducible diagrams in the presence
 of the background field.
If we set  $\tilde{\vf}=0$ in Eq.(\ref{test3}), we find that
\bea
\left. \tilde{\Gamma}[\tilde{\vf},B] \right|_{\tilde{\vf}=0}
&=& \left. \tilde{W}[J(B),B]-J\tilde{\vf} \right|_{\tilde{\vf}=0} \nn  \\
&=& \Gamma[B].
\eea 
This implies that the effective action $\Gamma[B]$ consists of
 one  $\tilde{\vf}$-particle irreducible bubble diagrams with no 
external $\tilde{\vf}$-lines.
Thus one can write the effective action as 
\be
e^{i\Gamma[B]}=\left. e^{i\tilde{W}[J,B]} \right|_{\tilde{\vf}=0}
=\left. N \int {\cal D}\f e^{iS[\f+B]+iJ\f} \right|_{\tilde{\vf}=0} .
\ee

To illustrate how the method facilitates the computation of the effective action,
we now consider the scalar $\f^4$ theory as an example.
The Lagrangian density of $\f^4$ theory can be written as 
\be
{\cal L}(\f)=-\ha \f(\partial^2+\m^2)\f -\fr{\l}{4!} \f^4 \equiv  -\ha \f (i\CD^{-1})\f-\fr{\l}{4!} \f^4,
\ee
where $D$ is the free propagator defined by
\be
\CD \equiv i(\partial^2+\m^2)^{-1}.
\ee
The shifted Lagrangian is then given by 
\be
{\cal L}(\f+B)=-i\ha BD^{-1}B- \fr {\l} {4!} B^4 -i(BD^{-1}-i\fr {\l}{3!}B^3)\f
-\ha\f iK^{-1}\f-\fr{\l}{3!}B\f^3-\fr{\l}{4!} \f^4,
\ee
where $K$ is the new propagator,
\be
K=i(\partial^2+\m^2+\fr{\l}{2}B^2)^{-1},
\label{kkk}
\ee
which includes a part of the interaction effect through the term proportional to $B^2$. 
The generating functional in the presence of the background field can
then be written as,
\be
e^{i\tilde{W}[J,B]}=N \det (-K)^{\ha} e^{\ha B \CD^{-1} B-\fr{i \l}{4!}B^4} 
e^{(B \CD^{-1} -\fr{i \l}{3!} B^3) \fr{\d}{i \d J}
 - \fr{i \l}{3!} B (\fr{\d}{i\d J})^3-\fr{i \l}{4!} (\fr{\d}{i\d J})^4}
 e^{\ha JKJ}.
\ee 

The effective action for the scalar $\f^4$ theory now becomes
\be
e^{i\Gamma[B]}=\left. e^{i\tilde{W}[J,B]} \right|_{\tilde{\vf}=0}
=N  e^{\ha B \CD^{-1} B-\fr{i \l}{4!}B^4} \det (-K)^{\ha} I_0[B],
\label{test4}
\ee
where
\be
I_0[B]=\left. e^{(B \CD^{-1} -\fr{i \l}{3!} B^3) \fr{\d}{i \d J}}
\exp [- \fr{i \l}{3!} B (\fr{\d}{i\d J})^3-\fr{i \l}{4!} (\fr{\d}{i\d J})^4 ]  
 e^{\ha JKJ} \right|_{\tilde{\vf}=0}.
\label{test5}
\ee
The coefficient of $I_0[B]$ in Eq.(\ref{test4}) represents the one-loop effective
action, and $I_0[B]$ represents the higher loop contributions to the effective
action. 
The first factor of Eq.(\ref{test5}) constitutes a shift operator, which 
does not change the extremum value of $\tilde{W}[J,B]$. 
Thus this factor, which
generates the tadpole diagrams, does not contribute to the effective action ,
and we can neglect it . 
One can therefore write $I_0[B]$ as   
\be
I_0[B]=\left. \exp [ - \fr{i \l}{3!} B (\fr{\d}{i\d J})^3-\fr{i \l}{4!} (\fr{\d}{i\d J})^4]  
 e^{\ha JKJ} \right|_{\tilde{\vf}=0}
\label{ioo}
\ee
which implies that the higher loop contributions to the effective action consist 
of one $\tilde{\varphi}$-irreducible bubble diagrams with no external  $\tilde{\varphi}$-lines.
One can compute the higher loop contributions by the Feynman rules:

\vspace{1cm}
\special{em:linewidth 0.4pt}
\unitlength 0.8mm
\linethickness{0.4pt}
\begin{picture}(110.00,70.00)
\emline{19.67}{70.00}{1}{40.00}{70.00}{2}
\emline{21.00}{60.00}{3}{30.00}{51.00}{4}
\emline{30.00}{51.00}{5}{39.00}{60.00}{6}
\emline{30.02}{50.91}{7}{30.02}{42.00}{8}
\emline{21.00}{37.00}{9}{39.33}{18.67}{10}
\emline{39.33}{37.00}{11}{20.67}{18.33}{12}
\put(29.97,51.12){\circle*{1.08}}
\put(30.18,27.87){\circle*{1.08}}
\put(50.00,70.00){\makebox(0,0)[lc]{$=K=\frac{i}{\partial^2+\mu^2+\frac{\lambda}{2}B^2}$}}
\put(50.00,51.33){\makebox(0,0)[lc]{$=-i \lambda B $}}
\put(49.67,27.33){\makebox(0,0)[lc]{$=-i \lambda $}}
\put(95.00,70.00){\makebox(0,0)[lc]{: propagator}}
\put(95.00,51.33){\makebox(0,0)[lc]{: 3-vertex}}
\put(95.00,27.33){\makebox(0,0)[lc]{: 4-vertex}}
\end{picture}

One can easily show that the two-loop contribution to the effective action 
consists of two bubble diagrams with no external lines\cite{r17}:
\bea
\G^{(2)}[B]=
\begin{array}{l}
\epsfig{figure=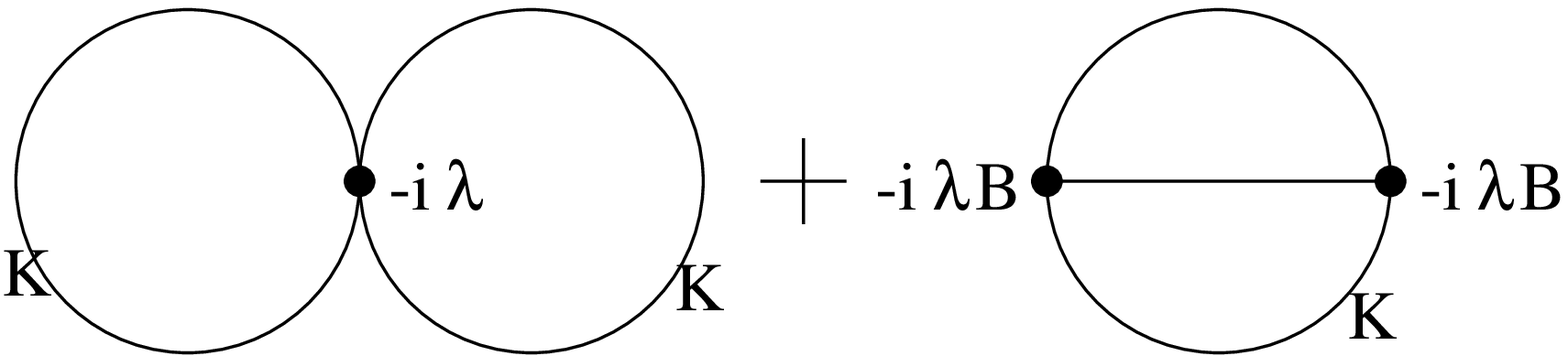,height=1.7cm}
\end{array}.
\label{fig111}
\eea
 Comparing this with the conventional method of computing the effective action, one 
can easily convince oneself that the background field method provides a very simple and effective 
way to compute the effective action in the loop expansion.

\subsection{Expansion around the Gaussian effective action}

 The reason why the procedure of computing the  effective action in 
the background field method is simplified is that the propagator used in the Feynman 
rules, Eq.(\ref{kkk}), includes a part of the interaction effect through the background 
field $B(x)$, and the generating functional has been rearranged in such a way that 
the zeroth order term in the expansion, the coefficient of $I_0 [B]$ in Eq.(\ref{test4}), is 
the one-loop effective action.

 We now want to proceed one step further and rearrange the effective action in such 
a way that the propagator used in the  expansion would include the non-perturbative 
effect and the zeroth order contribution would be the result of the non-perturbative 
approximation, such as the result of the Gaussian approximation.

Note that the Gaussian approximation consists of the cactus type diagrams where an 
internal line comes out of a point in space-time and goes back to the same point,
such as the first term of Eq.(\ref{fig111}).
 So we need to extract such diagrams out of the $I_0 [B]$ part of the effective action. 
To do this we examine the functional derivatives of the generating functional $e^{\fr{1}{2}JGJ}$ 
with respect to $J(x)$ :
\bea
\fr{\d}{\d J_x} e^{\ha J_y G_{yz} J_z} &=&
G_{xy} J_y e^{\ha J_y G_{yz} J_z} \equiv
(GJ)_x e^{\ha JGJ} \\
\fr{\d^2}{\d J_x^2} e^{\ha J_y G_{yz} J_z} &=&
((GJ)_x^2+G_{xx}) e^{\ha JGJ} \\
\fr{\d^3}{\d J_x^3} e^{\ha J_y G_{yz} J_z} &=&
((GJ)_x^3+3G_{xx} (GJ)_x) e^{\ha JGJ}
\label{j3} \\
\fr{\d^4}{\d J_x^4} e^{\ha J_y G_{yz} J_z} &=&
((GJ)_x^4+6G_{xx} (GJ)_x^2+3G_{xx}^2) e^{\ha JGJ} ,
\label{j4}
\eea
which would appear in the expansion of $I_0 [B]$.
Note that these derivatives, which generates $2$-loop and higher order 
contributions 
to $\Gamma[B]$, generate the cactus type diagrams as can be seen in Eqs.(\ref{j3}) and (\ref{j4}).

 To extract such diagrams out of $I_0 [B]$, we define new functional derivatives 
so that the following relations satisfy:
\bea
(\fr{\d^3}{\d J_x^3})' e^{\ha JGJ} &=&  (GJ)_x^3 e^{\ha JGJ}
\label{jp3}\\
(\fr{\d^4}{\d J_x^4})'  e^{\ha JGJ} &=& (GJ)_x^4 e^{\ha JGJ}.
\label{jp4}
\eea
 Comparing Eqs.(\ref{j3})-(\ref{j4}) and Eqs.(\ref{jp3})-(\ref{jp4}), we find that the primed derivatives 
must be defined by
\bea
(\fr{\d^3}{\d J_x^3})' & \equiv &
\fr{\d^3}{\d J_x^3}-3 G_{xx} \fr{\d}{\d J_x} 
\label{last3} \\
(\fr{\d^4}{\d J_x^4})' & \equiv &
\fr{\d^4}{\d J_x^3}-6 G_{xx} \fr{\d^2}{\d J_x^2}+3G_{xx}^2.
\label{last4}
\eea
 Note that the primed derivatives operated on the functional $e^{\fr{1}{2}JGJ}$ do 
not generate the cactus type diagrams that contribute to the Gaussian effective action.
This implies that by using the primed derivatives one may
in fact rewrite $I_0 [B]$ in such a way that all the cactus type diagrams would be included
in the zeroth order term in the expansion.

 To do this we need to find a new Green's function $G$ such that
\be
e^{\fr{\l B}{3!}(\fr{\d}{\d J})^3
-\fr{i \l}{4!}(\fr{\d}{\d J})^4} e^{\ha JKJ}= 
 N' e^{A \fr{\d}{\d J}} e^{\fr{\l B}{3!}(\fr{\d^3}{\d J^3})'
  -\fr{i\l}{4!}(\fr{\d^4}{\d J^4})'} e^{\ha JGJ},
\label{dool}
\ee
 where a normalization constant $N'$ and a function $A$ are to be determined. Using the 
definitions (\ref{last3}) and (\ref{last4}), one easily finds $G$, $N'$ and $A$ that satisfy Eq.(\ref{dool}):
\bea
N' &=&  \det (-G)^{\ha} \det (-G^{-1}+\fr{i \l}{2} G_{xx})^{\ha}
e^{\fr{i\l}{8}G_{xx}^2},  \\
A &=& \fr{\l}{2} B G_{xx}, \\
G_{xy}^{-1} &=& K_{xy}^{-1}+\fr{i\l}{2}G_{xx} \d_{xy}  \nn \\
&=& \CD_{xy}^{-1}-\fr{i \l}{2}B_x^2 \d_{xy}+\fr{i\l}{2}G_{xx} \d_{xy}.
\label{pregap}
\eea
 Using these results we can write the generating functional in the presence of the 
background field as 
\bea
e^{i \tilde{W}[J,B]} &=& \det (-G)^{\ha} e^{\ha B\CD^{-1}B
-\fr{i\l}{4!}B^4 +\fr{i\l}{8}G_{xx}^2}  \nn \\ 
&\times& e^{(-iB \CD^{-1}-\fr{\l}{3!}B^3 
+\fr{\l}{2} B G_{xx}) \fr{\d}{\d J}} 
e^{\fr{\l B}{3!}(\fr{\d^3}{\d J_x^3})'
  -\fr{i\l}{4!} (\fr{\d^4}{\d J_x^4})'}
e^{\ha JGJ}.
\eea

 The effective action can thus be written as
\bea
e^{i \G[B]} &=& \left. e^{\tilde{W}[J,B]} \right|_{\tildef=0} \nn \\
&=& \det (-G)^\ha e^{\ha B\CD^{-1}B
-\fr{i\l}{4!}B^4 +\fr{i\l}{8}G_{xx}^2} \; I[B],
\label{effec}
\eea
where
\bea
I[B] &=& \exp \left[ (-iB \CD^{-1}-\fr{\l}{3!}B^3 
+\fr{\l}{2} B G_{xx}) \fr{\d}{\d J} \right]  \nn \\
& &\times \left. \exp[\fr{\l B}{3!}(\fr{\d^3}{\d J_x^3})'
  -\fr{i\l}{4!} (\fr{\d^4}{\d J_x^4})'] e^{\ha JGJ} \right|_{\tildef=0} \nn \\
&=& \left. \exp[\fr{\l B}{3!}(\fr{\d^3}{\d J_x^3})'
  -\fr{i\l}{4!} (\fr{\d^4}{\d J_x^4})'] e^{\ha JGJ} \right|_{\tildef=0}.
\label{ib}
\eea
 In Eq.(\ref{ib}) we have used the fact that the shift operator, the tadpole part of the 
first line, does not affect the extremum value.
 Note that Eq.(\ref{pregap}), which determines the Green's function $G_{xy}$, is exactly the variational 
gap equation written in the Minkowski space notation, and therefore the coefficient of 
$I[B]$ in Eq.(\ref{effec}) gives rise to the Gaussian effective action as we have alluded to above.
 Since $I[B]$ can be expanded as a power series in $\lambda$ as is done in the conventional 
background field method, we have the perturbative expansion of the effective action around 
the Gaussian approximation.

 The structure of $I[B]$ is the same as that of $I_0 [B]$ in Eq.(\ref{ioo}), except that the 
propagator is now replaced by $G_{xy}$ and the functional derivatives are replaced by the primed 
derivatives. 
Thus the $I[B]$ contribution to the effective action 
consists of one $\tilde{\varphi}$-particle irreducible bubble 
diagrams with no external $\tilde{\varphi}$-lines, and without the cactus type diagrams.
 The contributions from $I[B]$ can therefore be evaluated by the Feynman rules:

\vspace{2cm}
\special{em:linewidth 0.4pt}
\unitlength 0.80mm
\linethickness{0.4pt}
\begin{picture}(95.00,71.01)
\emline{19.67}{70.00}{1}{40.00}{70.00}{2}
\emline{21.00}{60.00}{3}{30.00}{51.00}{4}
\emline{30.00}{51.00}{5}{39.00}{60.00}{6}
\emline{21.00}{37.00}{7}{39.33}{18.67}{8}
\emline{39.33}{37.00}{9}{20.67}{18.33}{10}
\put(50.00,70.00){\makebox(0,0)[lc]
{$=K=\frac{i}{\partial^2+\mu^2+\frac{\lambda}{2}B^2-\frac{\lambda}{2}G_{xx}}$}}
\put(50.00,51.33){\makebox(0,0)[lc]{$=-i \lambda B $}}
\put(49.67,27.33){\makebox(0,0)[lc]{$=-i \lambda $}}
\put(100.00,70.00){\makebox(0,0)[lc]{: propagator}}
\put(100.00,51.33){\makebox(0,0)[lc]{: 3-vertex}}
\put(100.00,27.33){\makebox(0,0)[lc]{: 4-vertex}}
\emline{19.67}{70.67}{11}{40.00}{70.67}{12}
\put(19.63,70.33){\circle*{1.29}}
\put(39.98,70.33){\circle*{1.36}}
\emline{21.74}{60.48}{13}{29.95}{52.26}{14}
\emline{29.95}{52.26}{15}{38.41}{60.71}{16}
\emline{30.48}{51.38}{17}{30.48}{41.93}{18}
\emline{29.48}{51.50}{19}{29.48}{41.99}{20}
\put(29.95,51.68){\circle*{1.68}}
\emline{20.15}{36.44}{21}{38.74}{17.93}{22}
\emline{40.00}{36.44}{23}{21.26}{17.63}{24}
\put(30.15,27.19){\circle*{1.57}}
\end{picture}

We have thus established the systematic method of computing the order-by-order correction 
terms to the Gaussian effective action.

 One can easily show that the ${\lambda}^2$ order contribution of $I[B]$, for example, consists 
of the diagrams,
\bea
I^{(2)}[B]=
\begin{array}{l}
\epsfig{figure=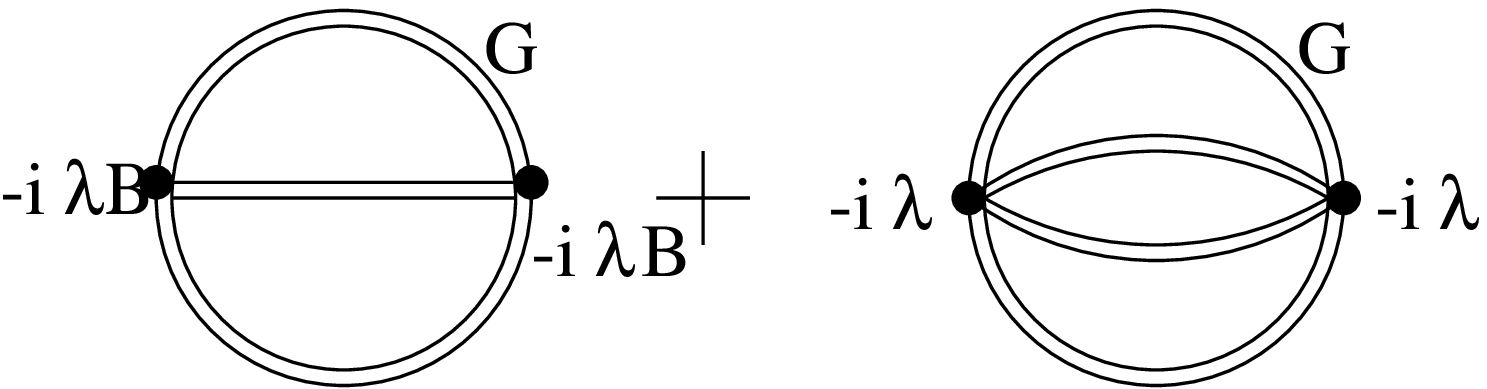,height=1.7cm} 
\end{array}.
\label{figxxx}
\eea
 Note that the first diagram of Eq.(\ref{fig111}) is a cactus type diagram and does not appear in 
the expansion of $I[B]$.

\subsection{Effective potential}
If we consider the space-time independent background field in the above formulation, we 
obtain the effective potential defined by
\be
V_{eff}(B) \equiv -\fr{\G[B]}{\int d^nx}.
\ee
 Since $G_{xy}$ is a function of $(x-y)$ for a constant $B$, one can solve Eq.(\ref{pregap}) by 
Fourier transformation,
\be
G(x-y) \equiv \int \fr{d^np}{(2\pi)^n} g(p) e^{ip(x-y)}.
\label{furi}
\ee
The Green's function $G_{xy}$ in the momentum space can then be written as
\be
g(p)=\fr{1}{i(p^2-m_G^2)},
\label{gpis}
\ee
where
\be
m_G^2 \equiv \m^2+\fr{\l}{2}B^2-\fr{\l}{2} \int \fr{d^np}{(2\pi)^n} g(p).
\label{mgis}
\ee
 Note that Eq.(\ref{gpis}) is the variational gap equation of Gaussian approximation written in 
the Minkowski notation.

 Up to the ${\lambda}^2$ order contribution of $I[B]$, the effective potential is therefore 
given by
\be
V_{eff}(B) = V_G+V_P, 
\ee
where
\be
V_G \equiv  \fr{\m^2}{2} B^2+\fr{\l}{4!}B^4
-\fr{1}{2\l}(m^2-\m^2-\fr{\l}{2}B^2)^2
+i \tr \log \sqrt{-G},
\label{vgis} 
\ee
and
\be
V_P  \equiv  i \fr{ \l^2}{12}B^2 G^3 -i \fr{\l^2}{48} G^4.
\label{vpis}
\ee
 $V_G$ of (\ref{vgis}) is the Gaussian effective potential and $V_P$ of (\ref{vpis}) is the first non-trivial 
contribution from $I[B]$, which is represented by the Feynman diagrams shown in (\ref{figxxx}). This 
result is the same as that of Cea and Tedesco\cite{r3} computed from the variational perturbation 
theory mentioned earlier.

 To illustrate the effectiveness of the perturbative correction method, we now consider the 
$(0+1)$-dimensional case, which is the quantum mechanical anharmonic oscillator. It is 
convenient to write the Green's function in Euclidean space which is defined by
\be
p^2=-p_E^2, \;\;\; d^np=id^n p_E.
\ee
Then $G(x,x)$ in Eq.(\ref{mgis}) can be written as
\be
\int \fr{dp}{2\pi} \fr{1}{i(p^2-m^2)}=
-\int \fr{dp_E}{2\pi} \fr{1}{p_E^2+m^2}=-\fr{1}{\sqrt{2}} \fr{1}{\sqrt{m^2}},
\ee
and the gap equation becomes
\be
m^2 = \m^2+\fr{\l}{2} B^2 + \fr{\l}{4 \sqrt{m^2}}.
\label{msgap}
\ee
 The effective potential up to the ${\lambda}^2$ order of $I[B]$ becomes
\be
V_{eff}(B) = V_G+V_P ,
\label{pot.gen} 
\ee
where
\be
V_G = \fr{\m^2}{2} B^2+\fr{\l}{4!}B^4 
-\fr{1}{2\l}(m^2-\m^2-\fr{\l}{2}B^2)^2+ \fr{\sqrt{m^2}}{2},
\label{pot.gauss} 
\ee
and
\be
V_P = -\fr{\l^2 }{144 m^4} B^2 -
\fr{\l^2}{1536 (\sqrt{m^2})^5}.
\label{pot.pert}
\ee

We can compare this result with other approximation methods. 
For the comparison of the 
effectiveness of various approximation methods 
it is convenient to compare the ground state energy eigenvalue of the anharmonic 
oscillator, which is defined by the minimum value of $V_{eff}$. 
The table below shows the 
errors of various approximation methods compared to the numerical results\cite{r18}. 
\begin{table}[h]
\begin{center}
\caption{Errors of $E_0 (\m^2=1)$ compared to the numerical result \label{ground}}
\begin{tabular}{|c|r|r|r|r|}
\hline \hline

$\l$&
1-loop & 2-loop &
Gaussian Approx. & Improved Gaussian \\  \hline

0.24 &
1.4\%&      0.05\% &   0.006\% & 0.0003\%  \\

24 &
38\%&      56\%&
1.1\%&     0.37\%  \\
\hline \hline
\end{tabular}
\end{center}
\end{table}
 This shows that our method greatly improves the Gaussian approximation even at the first 
non-trivial correction level, for all values of the coupling constant. Note that, for strong 
coupling, the Gaussian and the improved Gaussian approximation display much better results 
than the loop expansion methods, which shows the non-perturbative nature of the approximation 
methods.

\section {Conclusion}

 We have shown that the functional Schr\"odinger picture formulation of quantum field theory 
provides a conceptually simple non-perturbative approximation method, namely the Gaussian approximation.
 It has been proved to be a reliable method to obtain non-perturbative informations on the 
symmetry breaking phenomena, existence of bound states and time-evolution of the time-dependent systems 
with relatively little computational difficulties. The method, however, has a serious drawback in 
that it does not provide a systematic procedure to improve the approximation.

 We have shown that the background field method provides a solution to this limitation of the 
variational method. 
We were able to rearrange the effective action in such a way that the zeroth order contribution 
in the expansion of the effective action consists of the Gaussian effective action. This provides 
a new systematic expansion method which enables one to compute the order-by-order correction 
terms to the Gaussian effective action.

 The effective potentials for the $(0+1)$, $(1+1)$ and $(2+1)$ dimensional ${\phi}^4$ theory 
computed by the new method agree with those of the variational perturbation theory. It has been 
shown that the perturbative correction greatly improves the Gaussian approximation even at 
the first non-trivial correction level. 
Our method can also be applied to the time-dependent systems and can easily be extended to 
the cases of fermionic and gauge field theories.

 Although the effective action expanded around the Gaussian approximation for the 
lower-dimensional ${\phi}^4$ theories can easily be shown to be 
renormalizable, one has to study the renormalization procedure systematically for the $(3+1)$ dimensional
case.

\newpage
\begin{center}
{\Large \bf Acknowledgements}
\end{center}
This work was supported in part by the Korea Science and Engineering
Foundation under Grant 
No.95-0701-04-01-3 and 065-0200-001-2,
the Center for Theoretical Physics (SNU), and 
the Basic Science Research Institute Program, Ministy of Education,
under project No. BSRI-96-2425.


\eject

\end{document}